\documentclass[aps,pra,onecolumn,showpacs,floatfix]{revtex4}
\usepackage{graphicx}
\usepackage{amsmath}
\usepackage{bm}
\usepackage{bbm}

\usepackage{dcolumn}

\newcolumntype{.}{D{x}{}{-1}}

\newcommand{\bsigma}{\vec{\sigma}}

\newcommand{\bnabla}{\vec{\nabla}}

\newcommand{\bfr}{\vec{r}}

\newcommand{\bfp}{\vec{p}}

\newcommand{\Za}{{Z\alpha}}

\newcommand{\cE}{{\cal E}}

\newcommand{\lbr}{\left<} \newcommand{\rbr}{\right>}

\newcommand{\al}{\alpha}

\begin{document}

\title{Fine structure of helium and light helium-like ions}

\author{Krzysztof Pachucki} \affiliation{Institute of Theoretical
        Physics, University of Warsaw, Ho\.{z}a 69, 00--681 Warsaw, Poland}

\author{Vladimir A. Yerokhin} \affiliation{
Institute of Physics, University of Heidelberg, Philosophenweg
  12, D-69120 Heidelberg, Germany
and
Gesellschaft f\"ur Schwerionenforschung, Planckstra{\ss}e 1,
D-64291 Darmstadt, Germany
and
Center for Advanced Studies,
St.~Petersburg State Polytechnical University, Polytekhnicheskaya 29,
St.~Petersburg 195251, Russia }

\begin{abstract}
Calculational results are presented for the fine-structure splitting of the $2^3P$
state of helium and helium-like ions with the nuclear charge $Z$ up to
10. Theoretical predictions are in agreement with the latest experimental
results for the helium fine-structure intervals as well as with the most of
the experimental data available for light helium-like ions.
Comparing the theoretical value of the $2^3P_0-2^3P_1$ interval in
helium  with the experimental result [T.~Zelevinsky {\em et al.}
 Phys. Rev. Lett. {\bf 95}, 203001 (2005)], we
determine the value of the fine-structure constant $\alpha$ with an accuracy
of 31 parts per billion.
\end{abstract}


\maketitle

\section{Introduction}

The fine structure splitting of the $2^3P$ state in helium plays a special
role in atomic spectroscopy because it can be used for an accurate
determination of the fine structure constant $\alpha$. This fact was first
pointed out by Schwartz in 1964 \cite{schwartz:64}. The attractive features of
the fine structure splitting in helium as compared to other atomic transitions
are, first, the long lifetime of the metastable $2^3P_J$ levels (roughly two
orders of magnitude larger than that of the $2p$ state in hydrogen) and,
second, the relative simplicity of the theory of the fine structure.
Schwartz's suggestion stimulated a sequence of calculations
\cite{douglas:74,hambro:72,hambro:72:a,hambro:73}, which resulted in a
theoretical description of the helium fine structure complete up to order
$m\alpha^6$ (or $\alpha^4$~Ry) and a value of $\alpha$ accurate to
0.9~ppm \cite{lewis:78}.

The present experimental precision for the fine-structure intervals in helium is
sufficient for a determination of $\alpha$ with an accuracy of 14~ppb
\cite{zelevinsky:05,borbely:09}. In order to match this level of accuracy in
the theoretical description of the fine structure, the complete calculation of the
next-order, $m\alpha^7$ contribution and an estimation of the higher-order
effects is needed. The work towards this end started in 1990s and
extended over two decades
\cite{yan:95:prl,zhang:96:a,zhang:96:b,zhang:96:prl,zhang:96:c,zhang:97,pachucki:99:jpb,%
pachucki:00:jpb,pachucki:02:jpb:a,drake:02:cjp,pachucki:03:jpb}. In 2006, the
first complete evaluation of the $m\alpha^7$ correction to the helium fine
structure was reported \cite{pachucki:06:prl:he}. However, the
numerical results  presented there
disagreed with the experimental values by more than 10 standard
deviations.

In our recent investigations \cite{pachucki:09:hefs,pachucki:10:hefs}, we
recalculated all effects up to order $m\alpha^7$ to the fine structure of helium
and performed calculations for helium-like ions with nuclear
charges $Z$ up to 10. The calculations were extensively checked by studying
the hydrogenic ($Z\to\infty$) limit of individual corrections
and by comparing them with the results known from the hydrogen theory.
We found several problems in previous studies. As a result,
the present theoretical predictions are in agreement with the latest
experimental data for the fine-structure intervals in helium, as well as with
the most of experimental data available for light helium-like ions.
Comparison of our theoretical prediction for the $2^3P_0-2^3P_1$ interval in
helium (accurate to
57~ppb) with the experimental value \cite{zelevinsky:05} (accurate to
24~ppb) determines the value of the fine structure constant
$\alpha$ with an accuracy
of 31~ppb.

The calculation of the $m\alpha^7$ correction for the fine-structure splitting
of light helium-like atoms was reported in our recent Letter
\cite{pachucki:10:hefs}. In this paper, we present an extended description of
the $m\alpha^7$ correction
and a detailed term-by-term
comparison of our results with independent calculations by
Drake~\cite{drake:02:cjp} for helium and by Zhang {\em et al.}
\cite{zhang:96:prl} for helium-like ions.

\section{The spin-dependent $\bm{m}\bm{\alpha}^{\bm{7}}$ correction}

The $m\alpha^7$ correction to the fine-structure splitting of a two-electron atom
can be conveniently separated into four parts,
\begin{align} \label{ma7}
E^{(7)} \equiv m\alpha^7 \cE^{(7)}
= m\alpha^7\biggl[\cE^{(7)}_{\rm log}+ \cE^{(7)}_{\rm first} + \cE^{(7)}_{\rm
    sec}+ \cE^{(7)}_{L} \biggr]\,.
\end{align}
The first term in the brackets above combines all terms with $\ln Z$ and $\ln \alpha$
\cite{zhang:96:a,zhang:96:b,zhang:96:prl,pachucki:99:jpb,pachucki:06:prl:he},
\begin{align} \label{E7log}
\cE^{(7)}_{\rm log} &\ =
\ln[(Z\,\alpha)^{-2}] \,\left[
\lbr \frac{2 Z}{3}\,
i\,\bfp_1\times\delta^3(r_1)\,\bfp_1\cdot\bsigma_1 \rbr
  \right.
   \nonumber \\ &
-\lbr \frac{1}{4}
   (\bsigma_1\cdot\bnabla)\,(\bsigma_2\cdot\bnabla) \delta^3(r)\rbr
-\lbr \frac{3}{2}\,
i\,\bfp_1\times\delta^3(r)\,\bfp_1\cdot\bsigma_1\rbr
   \nonumber \\ &
   \left.
 + \frac{8Z}{3} \lbr H^{(4)}_{\rm fs} \frac1{(E_0-H_0)'} \bigl[
 \delta^3(r_1)+\delta^3(r_2)\bigr]\rbr
\right] \,,
\end{align}
where $\vec{r} = \vec{r}_1-\vec{r}_2$,
$H_0$ and $E_0$ are the Schr\"odinger Hamiltonian and its eigenvalue,
and $H^{(4)}_{\rm fs}$ is the spin-dependent part of the Breit-Pauli
Hamiltonian,
\begin{align}
H^{(4)}_{\rm fs}   = \ &
\frac{1}{4 \,r^3}\biggl[
\bigl(\bsigma_2+2\,\bsigma_1\bigr)\cdot\bfr\times\bfp_2
-\bigl(\bsigma_1+2\,\bsigma_2\bigr)\cdot\bfr
\times\bfp_1\biggr]
    \nonumber \\
+ &   \frac{Z}{4}
\left(
\frac{\bfr_1}{r_1^3}\,\times\bfp_1\cdot\bsigma_1+
\frac{\bfr_2}{r_2^3}\,\times\bfp_2\cdot\bsigma_2
\right)
\nonumber \\
+ &
\frac{1}{4}\left(
\frac{\bsigma_1\cdot\bsigma_2}{r^3}
-3\,\frac{\bsigma_1\cdot\bfr\,
\bsigma_2\cdot\bfr}{r^5}\right)\,
\,.
\end{align}

The second part of $\cE^{(7)}$
is induced by effective Hamiltonians to order $m\alpha^7$.
They were derived by one of us
(K.P.) in Refs. \cite{pachucki:06:prl:he,pachucki:09:hefs}.
(The previous derivation of this correction
by Zhang \cite{zhang:96:a,zhang:96:b} turned out to be not entirely consistent.)
The result is
\begin{align}
\cE^{(7)}_{\rm first} = \Bigl<  H_Q + H_H + H^{(7)}_{\rm fs, amm} \Bigr>\,.
\end{align}
The Hamiltonian $H_Q$ is induced by the two-photon exchange
between the electrons, the electron self-energy and the vacuum
polarization. It is given by
\cite{pachucki:06:prl:he}
\begin{align} \label{HQ}
H_Q  &\ = Z\,\frac{91}{180}
\,i\,\bfp_1\times\delta^3(r_1)\,\bfp_1\cdot\bsigma_1
  \nonumber \\ &
-\frac12\, (\bsigma_1\cdot\bnabla)\,(\bsigma_2\cdot\bnabla)\, \delta^3(r)\,
 \left[ \frac{83}{30}+\ln Z\right]
 \nonumber \\ &
+3\,i\,\bfp_1\times\delta^3(r)\,\bfp_1\cdot\bsigma_1\,
  \left[ \frac{23}{10}-\ln Z\right]
 \nonumber \\ &
-\frac{15}{8\,\pi}\, \frac{1}{r^7}\,(\bsigma_1\cdot\bfr)\,(\bsigma_2\cdot\bfr)
-\frac{3}{4\,\pi}\,i\,\bfp_1\times \frac{1}{r^3}\, \bfp_1\cdot\bsigma_1\,.
\end{align}
Here, the terms with $\ln Z$ compensate
the logarithmic dependence implicitly present in
expectation values of singular operators $1/r^3$ and $1/r^5$, so that
matrix elements of $H_Q$ do not have any logarithms in their $1/Z$ expansion.
The singular operators are defined though their integrals with
the arbitrary smooth function $f$,
\begin{align} \label{def1}
\int d^3 r \frac{1}{r^3}\,f(\bfr) \equiv \lim_{\epsilon\rightarrow 0} \int
\,& d^3 r\,\biggl[\frac{1}{r^3}\,\theta(r-\epsilon)
\nonumber \\ &
+ 4\,\pi\,\delta^3(r)\,(\gamma+\ln\epsilon)\biggr]\,f(\bfr) \,,
\end{align}
and
\begin{align} \label{def2}
\int d^3 r\,\frac{1}{r^7}&\,\biggl(r^i\,r^j-\frac{\delta^{ij}}{3}\,r^2\biggr)\,f(\bfr) \equiv
\nonumber \\ &
\lim_{\epsilon\rightarrow 0} \int
d^3 r\,
\biggl[\frac{1}{r^7}\,\biggl(r^i\,r^j-\frac{\delta^{ij}}{3}\,r^2\biggr)\theta(r-\epsilon)
\nonumber \\ &
+ \frac{4\,\pi}{15}\,\delta^3(r)\,(\gamma+\ln\epsilon)\,
\biggl(\partial^i\,\partial^j-\frac{\delta^{ij}}{3}\,\partial^2\biggr)\biggr]\,f(\bfr)\,,
\end{align}
where $\gamma$ is the Euler constant.

The effective Hamiltonian $H_H$ represents the anomalous magnetic moment (amm) correction
to the Douglas-Kroll $m\alpha^6$ operators and is given by
\cite{pachucki:06:prl:he}
\begin{align} \label{EH}
H_H &\ = -\frac{Z}{4}\,p_1^2\,\frac{\bfr_1}{r_1^3}\times\bfp_1\cdot\bsigma_1
 -\frac{3\,Z}{4}\,\frac{\bfr_1}{r_1^3}\times\frac{\bfr}{r^3}\cdot\bsigma_1\,(\bfr\cdot\bfp_2)
 +\frac{3\,Z}{4}\,\frac{\bfr}{r^3}\cdot\bsigma_1\,\frac{\bfr_1}{r_1^3}\cdot\bsigma_2
 +\frac{1}{2\,r^4}\,\bfr\times\bfp_2\cdot\bsigma_1
 -\frac{3}{4\,r^6}\,\bfr\cdot\bsigma_1\,\bfr\cdot\bsigma_2
\nonumber \\&
- \frac{1}{4}\,p_1^2\,\frac{\bfr}{r^3}\times\bfp_1\cdot\bsigma_1
-\frac{1}{4}\,p_1^2\,\frac{\bfr}{r^3}\times\bfp_2\cdot\bsigma_1
-\frac{Z}{4\,r}\,\frac{\bfr_1}{r_1^3}\times\bfp_2\cdot\bsigma_1
-\frac{i}{2}\,p_1^2\,\frac{1}{r^3}\,\bfr\cdot\bfp_2\,\bfr\times\bfp_1\cdot\bsigma_1
+\frac{3\,i}{4\,r^5}\,\bfr\times(\bfr\cdot \bfp_2)\,\bfp_1\cdot \bsigma_1
\nonumber \\&
-\frac{3}{8\,r^5}\,\bfr\times(\bfr\times \bfp_1\cdot\bsigma_1)\,\bfp_2\cdot
\bsigma_2
-\frac{1}{8\,r^3}\,\bfp_1\cdot\bsigma_2\,\bfp_2\cdot\bsigma_1
+\frac{21}{16}\,p_1^2\,\frac{1}{r^5}\,\bfr\cdot\bsigma_1\,\bfr\cdot\bsigma_2
-\frac{3\,i}{8}\,p_1^2\,\frac{\bfr}{r^3}\cdot\bsigma_1\,\bfp_1\cdot\bsigma_2
\nonumber \\&
+\frac{i}{8}\,p_1^2\,\frac{1}{r^3}\,\biggl[\bfr\cdot\bsigma_2\,\bfp_2\cdot\bsigma_1
         +(\bfr\cdot\bsigma_1)\,(\bfp_2\cdot\bsigma_2)
         -\frac{3}{r^2}\,\bfr\cdot\bsigma_1\,\vec
          r\cdot\bsigma_2\,\bfr\cdot\bfp_2\biggr]
-\frac{1}{4}\,\bfp_1\cdot\bsigma_1\,\bfp_1\times\frac{\bfr}{r^3}\cdot\bfp_2
\nonumber \\&
+\frac{1}{8}\,\bfp_1\cdot\bsigma_1\,\biggl[-\bfp_1\cdot\bsigma_2\,\frac{1}{r^3}
          +3\bfp_1\cdot\bfr\,\frac{\bfr}{r^5}\cdot\bsigma_2\biggr]\,.
\end{align}
The Hamiltonian $H^{(7)}_{\rm fs}$ is the $m\alpha^7$ amm correction to the
Breit-Pauli Hamiltonian,
\begin{align}
H^{(7)}_{\rm fs}   = \ &
\frac{1}{2\pi^3 \,r^3}
\bigl(\bsigma_1+\bsigma_2\bigr)\cdot\bfr\times\bigl(\bfp_2-\bfp_1\bigr)\,c_3
    \nonumber \\
+ &   \frac{Z}{2\pi^3}
\left(
\frac{\bfr_1}{r_1^3}\,\times\bfp_1\cdot\bsigma_1+
\frac{\bfr_2}{r_2^3}\,\times\bfp_2\cdot\bsigma_2
\right)\,c_3
\nonumber \\
+ &
\frac{1}{2\pi^3}\left(
\frac{\bsigma_1\cdot\bsigma_2}{r^3}
-3\,\frac{\bsigma_1\cdot\bfr\,
\bsigma_2\cdot\bfr}{r^5}\right)
   \bigl( c_1c_2+c_3\bigr)\,
\,,
\end{align}
where $c_1 = 1/2$,
$c_2 = -0.328\,478\,965$ and $c_3 = 1.181\,241\,456$ are the expansion
coefficients of the free-electron amm in powers of $(\alpha/\pi)$.

The third part of $\cE^{(7)}$ is given by the second-order matrix elements
of the form
\cite{pachucki:06:prl:he}
\begin{align}
\cE^{(7)}_{\rm sec} =  &\
  2\lbr H^{(4)}_{\rm fs}\frac{1}{(E_0-H_0)'}H^{(5)}_{\rm nlog}\rbr
\nonumber \\ &
  + 2\lbr \biggl[ H^{(4)}_{\rm fs}+H^{(4)}_{\rm nfs}\biggr]
 \frac{1}{(E_0-H_0)'}H^{(5)}_{\rm fs}\rbr \,,
\end{align}
where $H^{(5)}_{\rm nlog}$ is the effective Hamiltonian responsible for the
nonlogarithmic $m\alpha^5$ correction to the energy,
\begin{align}
H^{(5)}_{\rm nlog} = - \frac{7}{6\,\pi\,r^3}
+ \frac{38 Z}{45}\, \left[\delta^3(r_1)+\delta^3(r_2)\right] \,,
\end{align}
$H^{(4)}_{\rm nfs}$ is the spin-independent part of the Breit-Pauli
Hamiltonian (with the term $\propto \delta^3(r)$ omitted since it does not
contribute in our case),
\begin{eqnarray} \label{nfs}
H^{(4)}_{\rm nfs} & = & -\frac{1}{8}\,(p_1^4+p_2^4)+
\frac{Z\,\pi}{2}\,\bigl[\delta^3(r_1)+\delta^3(r_2)\bigr]
\nonumber \\ & &
-\frac{1}{2}\,p_1^i\,
\biggl(\frac{\delta^{ij}}{r}+\frac{r^i\,r^j}{r^3}\biggr)\,p_2^j
\,,
\end{eqnarray}
and $H^{(5)}_{\rm fs}$ is the $m\alpha^5$ amm correction to $H^{(4)}_{\rm fs}$,
\begin{align}
H^{(5)}_{\rm fs}   = \ &
\frac{1}{4\pi \,r^3}
\bigl(\bsigma_1+\bsigma_2\bigr)\cdot\bfr\times\bigl(\bfp_2-\bfp_1\bigr)
    \nonumber \\
+ &   \frac{Z}{4\pi}
\left(
\frac{\bfr_1}{r_1^3}\,\times\bfp_1\cdot\bsigma_1+
\frac{\bfr_2}{r_2^3}\,\times\bfp_2\cdot\bsigma_2
\right)
\nonumber \\
+ &
\frac{1}{4\pi}\left(
\frac{\bsigma_1\cdot\bsigma_2}{r^3}
-3\,\frac{\bsigma_1\cdot\bfr\,
\bsigma_2\cdot\bfr}{r^5}\right)\,
\,.
\end{align}

The fourth part of $\cE^{(7)}$ is the
contribution induced by the
emission and reabsorption of virtual photons of low energy.
It is denoted as $\cE_{L}^{(7)}$ and interpreted as
the relativistic correction to the Bethe logarithm.
The expression for $\cE_{L}^{(7)}$ reads
\cite{pachucki:00:jpb}
\begin{align}
 \cE_{L}^{(7)} =
 -\frac{2}{3\,\pi}\,
\delta\, \Biggl< (\bfp_1+\bfp_2)\,\cdot (H_0-E_0)
\ln\left[\frac{2(H_0-E_0)}{Z^2}\right] &
(\bfp_1+\bfp_2)\Biggr>
  \nonumber \\ &
+\frac{i\,Z^2}{3\,\pi}
\Biggl< \left(\frac{\bfr_1}{r_1^3}+\frac{\bfr_2}{r_2^3}\right) \times
     \frac{\bsigma_1+\bsigma_2}{2}
\ln\left[\frac{2(H_0-E_0)}{Z^2}\right]
\left(\frac{\bfr_1}{r_1^3}+\frac{\bfr_2}{r_2^3}\right) \Biggr>\,,
\label{bethe}
\end{align}
where $\delta \lbr \ldots \rbr$ denotes the first-order perturbation of
the matrix element $\langle\ldots\rangle$ by $H^{(4)}_{\rm fs}$, implying
perturbations of the reference-state wave function, the reference-state
energy, and the electron Hamiltonian.

\section{Results: helium}

The summary of individual contributions to the fine-structure intervals
of helium is given in Table~\ref{tab:summary}. Numerical results are
presented for the large $\nu_{01}$ and the small $\nu_{12}$ intervals,
defined by
\begin{eqnarray}
\nu_{01} = \bigl[ E(2^3P_0)-E(2^3P_1)\bigr]/h\,, \\
\nu_{12} = \bigl[ E(2^3P_1)-E(2^3P_2)\bigr]/h\,.
\end{eqnarray}
We note that the style of
breaking the total result into separate entries used in
Table~\ref{tab:summary} differs from that used in the summary tables of
the previous papers by K.P. {\em et al.} \cite{pachucki:06:prl:he,pachucki:09:hefs}.
Particulary, the lower-order terms listed in Table III of
Ref.~\cite{pachucki:06:prl:he} and in Table~II of Ref.~\cite{pachucki:09:hefs}
contained contributions of higher orders, whereas in the present work,
the entries in Table~\ref{tab:summary} contain only the contributions of the
order specified.

\begin{table*}
\caption{
Summary of individual contributions to the fine-structure intervals in helium,
in kHz. The parameters \cite{mohr:08:rmp} are
$\alpha^{-1} = 137.035\,999\,679(94)$, $cR_{\infty}
= 3\,289\,841\,960\,361(22)$~kHz, and $m/M = 1.370\,933\,555\,70 \times 10^{-4}$.
The values by Drake are taken from Table 3 of Ref.~\cite{drake:02:cjp}.
The label $(+m/M)$ indicates that the corresponding entry comprises both
the non-recoil and recoil contributions of the specified order in $\alpha$.
The uncertainty due to the value of $\alpha$ is not shown. 
\label{tab:summary}
}
\begin{center}
  \begin{tabular}{l..c}
\hline
 Term &
\multicolumn{1}{c}{$\nu_{01}$}  &
        \multicolumn{1}{c}{$\nu_{12}$}  &    Ref.      \\
    \hline\\[-5pt]
$m\al^4(+m/M)$           &   29\,563\,765x.45        &        2\,320\,241x.43 &  \\
                         &   29\,563\,765x.23^a      &        2\,320\,241x.42^a & \cite{drake:02:cjp}  \\[0.1cm]
$m\al^5(+m/M)$           &        54\,704x.04        &         -22\,545x.00 &  \\
                         &        54\,704x.04        &         -22\,545x.01 & \cite{drake:02:cjp}  \\[0.1cm]
$m\al^6$                 &        -1\,607x.52(2)     &          -6\,506x.43 &  \\
                         &        -1\,607x.61(4)     &          -6\,506x.45(7)&  \cite{drake:02:cjp}  \\[0.1cm]
$m\al^6m/M$              &             -9x.96        &              9x.15 &  \\
                         &            -10x.37(5)     &              9x.80(11)&  \cite{drake:02:cjp}  \\[0.1cm]
$m\al^7 \log(Z\al)$      &             81x.43        &             -5x.87 &  \\
                         &             81x.42^b      &             -5x.87^b & \cite{drake:02:cjp}  \\[0.1cm]
$m\al^7$, nlog           &             18x.86        &            -14x.38 &  \\[0.1cm]
$m\al^8$                 &           \pm1x.7         &           \pm1x.7  &  \\[0.1cm]
Total theory             &   29\,616\,952x.29 \pm1.7 &        2\,291\,178x.91\pm1.7 &\\[0.1cm]
Experiment               &   29\,616\,951x.66(70)^c  &        2\,291\,177x.53(35)^f &\\[0.1cm]
                         &   29\,616\,952x.7(10)^d   &        2\,291\,175x.59(51)^c&\\[0.1cm]
                         &   29\,616\,950x.9(9)^e    &        2\,291\,175x.9(10)^g&\\
\hline
\end{tabular}
\end{center}

$^a$ the original result was scaled to the present value of $\al$.\\
$^b$ the original result was
altered by the substitution $\ln(\al)\to\ln(Z\al)$ in the terms proportional
to $\ln(\al)$, in order to comply with the present result for the
logarithmic $m\al^7$ contribution.\\
$^c$ Ref.~\cite{zelevinsky:05}.
$^d$ Ref.~\cite{giusfredi:05}.
$^e$ Ref.~\cite{george:01}.
$^f$ Ref.~\cite{borbely:09}.
$^g$ Ref.~\cite{castillega:00}.
\end{table*}

A term-by-term comparison with the independent calculation by Drake
\cite{drake:02:cjp} is made whenever possible. We observe good agreement
between the two calculations for the lower-order terms, namely, for the
$m\alpha^4$, $m\alpha^5$, and $m\alpha^6$
corrections. However, for the recoil correction to order $m\alpha^6$, our
results differ from Drake's ones by about 0.5~kHz for both intervals. The
reason for this disagreement seems to be different for the large and the small
intervals. For the large interval, the deviation is due to the recoil operator
part, whereas for the small interval, it is mainly due to the mass
polarization part (see discussion in Ref.~\cite{pachucki:09:hefs}).

Our estimates of the uncalculated higher-order effects for helium
are much larger than those in the previous studies
\cite{pachucki:02:jpb:a,drake:02:cjp}. The previous estimates amounted to
significantly less than 1~kHz for both intervals and were based on some
logarithmic contributions to order $m\alpha^8$ that were identified by
analogy with the hydrogen fine structure. We now believe that the dominant
$m\alpha^8$ contribution might be of relativistic origin. Our estimates of
$\pm$1.7~kHz for both intervals were obtained by multiplying the $m\alpha^6$
contribution for the sum of $\nu_{01}+\nu_{12}$ by the factor of $(\Za)^2$.

Our
result for the $\nu_{01}$ interval of helium agrees well with the experimental
values \cite{zelevinsky:05,giusfredi:05,george:01}. For the $\nu_{12}$ interval, our
theory is by about $2\sigma$ away from the values obtained in
Refs.~\cite{zelevinsky:05,castillega:00} but
in agreement with the latest measurement by Hessels and coworkers~\cite{borbely:09}.
Assuming the validity of the theory, we
combine the theoretical prediction
for the $\nu_{01}$ interval in helium with the
experimental result \cite{zelevinsky:05} and obtain
the following value of the fine structure constant,
\begin{equation} \label{alpha}
\alpha^{-1}({\rm He}) = 137.036\,001\,1\,(39)_{\rm theo}\,(16)_{\rm exp}\,,
\end{equation}
which is accurate to 31~ppb and agrees with the more precise results of
Refs.~\cite{hanneke:08,cadoret:08,jeffrey:97}.

\section{Results: helium-like ions}

\begin{table*}
\caption{Individual contributions
to the fine-structure intervals of helium-like atoms, in MHz$/Z^4$.
The label $(+m/M)$ indicates that the corresponding entry comprises both
the non-recoil and recoil contributions of the specified order in $\alpha$.
For $Z=3$, 7, and 10, a term-by-term comparison is made with the
previous calculation by Zhang {\em et. al.}~\cite{zhang:96:prl}. The results
of Ref.~\cite{zhang:96:prl} for the leading $m\alpha^4$ correction were scaled
to the present value of $\alpha$. The deviation
for the $m\alpha^7({\rm log})$ contribution is due to the difference in the
expressions for this term.
\label{tab:helike}}
\begin{center}
  \begin{tabular}{l.......c}
\hline\\[-0.3cm]
 $Z$ &
\multicolumn{1}{c}{$m\alpha^4(+m/M)$}
             &  \multicolumn{1}{c}{$m\alpha^5(+m/M)$}
                         &\multicolumn{1}{c}{$m\alpha^6$}
                               &\multicolumn{1}{c}{$m\alpha^6m/M$}
                                  &\multicolumn{1}{c}{$m\alpha^7({\rm log})$}
                                         &\multicolumn{1}{c}{$m\alpha^7({\rm nlog})$}
                                               &\multicolumn{1}{c}{\rm Total}
                                               &\multicolumn{1}{c}{\rm Ref.}   \\
    \hline\\[-5pt]
\multicolumn{6}{l}{  $\nu_{01}$}\\
  2 &  1847x.735\,34  &   3.41x9\,00 &  -0.10x0\,47 &  -0.0x00\,62 &  0.0x05\,09  &   0.0x01\,18 & 1851.x059\,52\,(11) \\
  3 &  1917x.793\,96  &   3.24x9\,78 &   1.23x0\,26 &  -0.0x02\,43 &  -0.0x10\,76 &   0.0x18\,01 & 1922.x278\,81\,(59) \\
    &  1917x.793\,97  &   3.24x9\,78 &   1.23x0\,25 &              &  -0.0x10\,27 &              & 1922.x26\,(2)&\cite{zhang:96:prl}\\
  4 &  1346x.965\,34  &   1.94x3\,84 &   4.56x6\,98 &  -0.0x06\,70 &  -0.0x28\,43 &   0.0x46\,48 & 1353.x4875\,(39) \\
  5 &   765x.885\,57  &   0.68x5\,51 &  10.37x4\,47 &  -0.0x14\,17 &  -0.0x41\,39 &   0.0x86\,28 & 776.x976\,(14) \\
  6 &   270x.387\,72  &  -0.36x7\,57 &  19.26x6\,47 &  -0.0x27\,89 &  -0.0x48\,86 &   0.1x39\,52 & 289.x349\,(37) \\
  7 &  -139x.085\,57  &  -1.22x9\,55 &  31.90x8\,79 &  -0.0x45\,30 &  -0.0x51\,10 &   0.2x09\,03 & -108.x294\,(83)\\
    &  -139x.085\,58  &  -1.22x9\,55 &  31.90x8\,82 &              &  -0.0x46\,33 &              & -108.x5\,(3)&\cite{zhang:96:prl}\\
  8 &  -477x.534\,46  &  -1.93x7\,91 &  48.98x8\,80 &  -0.0x68\,79 &  -0.0x48\,55 &   0.2x97\,85 & -430.x30\,(17) \\
  9 &  -759x.770\,39  &  -2.52x6\,32 &  71.20x3\,90 &  -0.0x93\,96 &  -0.0x41\,63 &   0.4x09\,16 & -690.x82\,(31) \\
 10 &  -997x.723\,26  &  -3.02x1\,03 &  99.25x7\,05 &  -0.1x37\,44 & -0.0x30\,76  &   0.5x46\,19 & -901.x11\,(53) \\
    &  -997x.723\,25  &  -3.02x1\,03 &  99.25x7\,05 &              & -0.0x21\,29  &              & -901.x5&\cite{zhang:96:prl}\\[0.1cm]
\multicolumn{6}{l}{
  $\nu_{02}$}\\
  2 &  1992x.750\,43 &   2.00x9\,94 &  -0.50x7\,12 &  -0.0x00\,05 &  0.0x04\,72  &   0.0x00\,28 & 1994.x258\,20\,(11) \\
  3 &  1150x.274\,90 &  -0.94x2\,85 &  -0.86x4\,60 &  -0.0x00\,05 &  -0.0x22\,16 &   0.0x14\,83 & 1148.x460\,07\,(41) \\
    &  1150x.274\,91  & -0.94x2\,85 &  -0.86x4\,60 &              &  -0.0x23\,48 &              & 1148.x44\,(2)&\cite{zhang:96:prl}\\
  4 &  -384x.659\,15 &  -4.44x8\,24 &  -1.38x9\,63 &  -0.0x00\,06 &  -0.0x45\,39 &   0.0x32\,04 & -390.x5104\,(12) \\
  5 & -1739x.328\,53 &  -7.32x0\,66 &  -2.39x3\,83 &  -0.0x00\,04 &  -0.0x54\,46 &   0.0x46\,61 & -1749.x0509\,(32) \\
  6 & -2838x.550\,28 &  -9.58x0\,33 &  -3.99x4\,54 &   0.0x00\,01 &  -0.0x48\,68 &   0.0x56\,88 & -2852.x1169\,(77) \\
  7 & -3724x.421\,92 & -11.37x0\,60 &  -6.24x5\,32 &   0.0x00\,08 &  -0.0x29\,03 &   0.0x62\,15 & -3742.x005\,(16) \\
    & -3724x.421\,93  &-11.37x0\,60 &  -6.26x3\,64 &              &  -0.0x41\,90 &              & -3742.x1\,(3)&\cite{zhang:96:prl}\\
  8 & -4445x.632\,74 & -12.81x2\,45 &  -9.17x4\,16 &   0.0x00\,17 &   0.0x03\,27 &   0.0x62\,07 & -4467.x554\,(31) \\
  9 & -5041x.009\,23 & -13.99x3\,89 & -12.79x7\,23 &   0.0x00\,25 &   0.0x47\,05 &   0.0x56\,47 & -5067.x697\,(55) \\
 10 & -5539x.338\,27 & -14.97x7\,37 & -17.12x4\,41 &   0.0x00\,38 &   0.1x01\,27 &   0.0x45\,23 & -5571.x293\,(91) \\
    & -5539x.338\,27  &-14.97x7\,38 & -17.14x5\,16 &              &   0.0x75\,67 &              & -5571.x4&\cite{zhang:96:prl}
\\[0.2cm]\hline
  \end{tabular}
\end{center}
\end{table*}

Table~\ref{tab:helike} gives the summary of individual contributions to the
fine-structure intervals of helium-like atoms with the nuclear charge number
$Z$ up to 10. We choose to present results for the intervals
$\nu_{01}$ and $\nu_{02}\equiv \nu_{01}+\nu_{12}$, and not for
$\nu_{01}$ and $\nu_{12}$, as is customary. The reason to consider
$\nu_{02}$ is that this interval is free from effects of the $2^3P_1-2^1P_1$
mixing, which strongly affect the $\nu_{01}$ and $\nu_{12}$ intervals.
As a result of the absence of the mixing effects, all corrections to $\nu_{02}$
starting with the order of $m\alpha^6$  demonstrate a weaker $Z$
dependence as compared to those to $\nu_{01}$ and $\nu_{12}$. The most drastic
difference occurs for the $\alpha^6m^2/M$ correction: for $Z=10$, this
correction for $\nu_{02}$ is by 3 orders of magnitude smaller than that for
$\nu_{01}$.

The uncertainty of the theoretical values specified in
Table~\ref{tab:helike} is solely due to uncalculated higher-order
effects. Its estimation for helium was already discussed. For helium-like
ions, we obtain the uncertainty by multiplying the $m\alpha^6$
contribution for the corresponding interval by the factor of $(\Za)^2$.
So, our error estimates are typically
by a factor of $1/Z$ smaller for the $\nu_{02}$ interval  than for the
$\nu_{01}$ (or, equivalently, $\nu_{12}$) interval.

For $Z=3$, 7, and 10, Table~\ref{tab:helike} presents a term-by-term
comparison  with the previous calculation by Zhang
{\em et. al.}~\cite{zhang:96:prl}.  We observe excellent agreement for the
$m\alpha^4$ and $m\alpha^5$ corrections. For the $m\alpha^6$ correction, the
agreement is excellent in all cases except for the $\nu_{02}$ interval and
$Z=7$ and 10, where a small deviation is present. The results of the two
calculations for the $m\alpha^7({\rm log})$ correction are different, but this is
explained by the difference in the expressions for this term. If we use the
same expression as in Ref.~\cite{zhang:96:prl}, excellent agreement is 
found again.

\begin{figure*}[thb]
  \centerline{\includegraphics[width=0.8\textwidth]{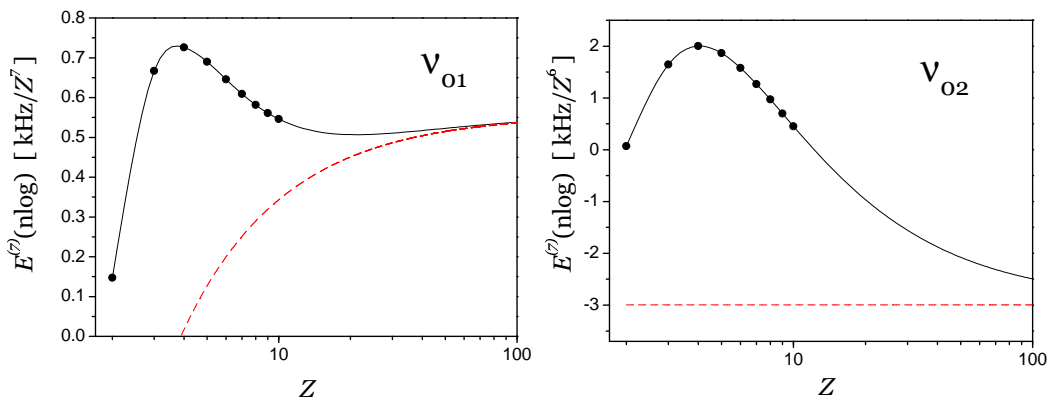}}
\caption{Nonlogarithmic $m\alpha^7$ correction to the fine-structure intervals
of helium-like atoms, for
the  $\nu_{01}$ interval (left) and for the $\nu_{02}$ interval (right).
Black dots denote the numerical results, solid line stands for the
fit of the $1/Z$ expansion, and the dashed (red) line indicates the asymptotic
high-$Z$ results. Note that the
results in different graphs are scaled by different factors. It is $Z^7$ for
the $\nu_{01}$ interval and $Z^6$, for $\nu_{02}$. The different $Z$ scaling is
the consequence of the triplet-singlet mixing effects.
  \label{fig:1} }
\end{figure*}

In Fig.~\ref{fig:1} we plot our numerical results for the $m\alpha^7$
correction as a function of the nuclear charge number $Z$, together with the
fit of the $1/Z$ expansion and with the asymptotical high-$Z$ limit of this
correction. The form of the $1/Z$ expansion and the values of the
first coefficient(s) are
known. For the $\nu_{02}$ interval, the leading term scales as $Z^6$ and is
calculated for hydrogen in Ref.~\cite{jentschura:96}. For the $\nu_{12}$
interval, there are additional $Z^7$ and $Z^6$ contributions due to the
triplet-singlet mixing, which are obtained by expanding the following
expression in $1/Z$,
$$
\delta E_{\rm mix} = \frac{\left|\langle
  2^1P_1|H^{(4)}_{\rm fs}|2^3P_1\rangle\right|^2}{E_0(2^3P_1)-E_0(2^1P_1)}\,.
$$
The resulting asymptotic form of the nonlogarithmic $m\alpha^7$ correction is
\begin{align}
\cE^{(7, {\rm nlog})}(\nu_{01})/Z^7 &\ = 0.004045 -0.015524/Z+ O(1/Z^2)\,,
 \\\
\cE^{(7, {\rm nlog})}(\nu_{02})/Z^6 &\ = -0.021706+ O(1/Z) \,.
\end{align}
By fitting the $1/Z$ expansion of our numerical data for $\cE^{(7, {\rm nlog})}$,
we were able to reproduce well the values of the coefficients given above,
which served as an important check of our calculations.

%
%

The comparison of the present theoretical predictions
with experiment data is summarized in Table~\ref{tab:compar}. The
agreement is very good in most cases. The only
significant discrepancy is for Be$^{2+}$, where the difference
of $1.7$ standard deviations ($\sigma$)
is observed for $\nu_{12}$ and that of $3.5\,\sigma$, for $\nu_{02}$.
It is important that for both the $\nu_{01}$ and $\nu_{12}$ intervals there
are experimental results available for helium-like ions, whose accuracy
significantly exceeds the theoretical errors. These are the
measurement of $\nu_{01}$ in helium-like nitrogen by Thompson {\em at
  al.}~\cite{thompson:98} and  that of $\nu_{12}$ in helium-like fluorine by
Myers {\em et al.}~\cite{myers:99:F}. Good agreement of the present theory
with these experimental results suggests that the theoretical errors (i.e.,
the uncalculated higher-order effects) were reasonably estimated.
It is unfortunate that there are no experimental results with comparable
accuracy available for the $\nu_{02}$ interval. Since $\nu_{02}$ is not
affected by the triplet-singlet mixing effects, accurate experimental results
for this interval in light helium-like ions
would yield an improved estimate for
the uncalculated higher-order effects in helium, thus increasing accuracy
of determination of $\alpha$ from the helium fine structure.

Comparing theoretical and experimental results for the fine structure of 
helium-like ions, one should keep in mind
that the present calculation is carried out for a spinless nucleus, whereas
the experimental results listed in Table~\ref{tab:compar} were performed for
non-zero nuclear spin isotopes. For a nucleus with spin, the hyperfine
splitting can usually be evaluated separately and employed for an
experimental determination of the fine structure. This procedure, however,
ignores the mixing between the hyperfine and the fine splittings. So, more accurate
calculations should account for both effects simultaneously.

In summary, the theory of the fine structure of helium and light helium-like
ions is now complete up to orders $m\alpha^7$ and $\alpha^6m^2/M$.
Theoretical predictions agree with the
latest experimental results for helium, as well as with most of the experimental
data for light helium-like ions. A combination of the theoretical and experimental results
for the $2^3P_0-2^3P_1$ interval in helium yields an independent determination of
the fine structure constant $\alpha$ accurate to 31~ppb.

\begin{table*}
\caption{Comparison of theoretical and experimental results for
the fine-structure intervals of helium-like ions.
Units are MHz for Li$^+$ and cm$^{-1}$ for other atoms.
\label{tab:compar}}
\begin{center}
  \begin{tabular}{l..r}
\hline\\[-0.3cm]
 $Z$ &
\multicolumn{1}{c}{Present theory}
             &  \multicolumn{1}{c}{Experiment}
                         &  Ref.        \\
    \hline\\[-5pt]
\multicolumn{4}{l}{$\nu_{01}$}\\
  3 &  155\,704.5x84(48)     &  155\,704.2x7(66)      & \cite{riis:94} \\
  4 &   11.5x57\,756(33)     &    11.5x58\,6(5)       & \cite{scholl:93} \\
  5 &   16.1x98\,21(29)      &    16.2x03(18)         & \cite{dinneen:91} \\
  7 &   -8.6x73\,1(67)       &    -8.6x70\,7(7)       & \cite{thompson:98} \\
  8 &  -58.7x91\,(23)        &   -59.2x\,(1.1)        & \cite{peacock:84}\\
 10 & -300.5x8(18)           &  -300.7x(2.2)          & \cite{peacock:84}
\\[5pt]
\multicolumn{4}{l}{$\nu_{12}$}\\
  9 &  -957.8x86(79)        &  -957.8x73\,0(12)      & \cite{myers:99:F}
\\[5pt]
\multicolumn{4}{l}{$\nu_{02}$}\\
  3 &    93\,025.2x66(34)   &  93\,025.8x6(61)       & \cite{riis:94} \\
  4 &     -3.3x34\,663(10)  &    -3.3x36\,4(5)       & \cite{scholl:93} \\
  5 &    -36.4x63\,787(66)  &   -36.4x57(16)         & \cite{dinneen:91} \\
  8 &   -610.3x92\,3(42)    &  -611.3x(7)            & \cite{peacock:84}\\
 10 &  -1858.3x83\,(30)     & -1858.3x(1.5)          & \cite{peacock:84}
\\[0.2cm]\hline
  \end{tabular}
\end{center}
\end{table*}

Support by NIST through Precision Measurement Grant PMG 60NANB7D6153
and by the Helmholtz Gemeinschaft
(Nachwuchsgruppe VH-NG-421) is gratefully acknowledged.


\end{document}